\documentclass[12pt,preprint]{aastex}

\shorttitle{Strong activity in \twom}
\shortauthors{Reiners et al.}

\begin{document}

\newcommand{\twom}{2MASS\,0535$-$0546}


\title{Detection of strong activity in the eclipsing binary brown
  dwarf 2MASS~J05352184$-$0546085 -- A possible explanation for the
  temperature reversal}


\author{A. Reiners\altaffilmark{*}, A. Seifahrt}
\affil{Institut f\"ur Astrophysik, Georg-August-Universit\"at, Friedrich-Hund-Platz 1, D-37077 G\"ottingen}
\email{[Ansgar.Reiners, seifahrt]@phys.uni-goettingen.de}
\author{K.G. Stassun}
\affil{Department of Physics and Astronomy, Vanderbilt University, Nashville, TN 37235}
\email{keivan.stassun@vanderbilt.edu}
\author{C. Melo}
\affil{European Southern Observatory, Casilla 19001, Santiago 19, Chile}
\email{cmelo@eso.org}
\and
\author{R.D. Mathieu}
\affil{Department of Astronomy, University of Wisconsin -- Madison, Madison, WI 53706}
\email{mathieu@astro.wisc.edu>}


\altaffiltext{*}{Emmy Noether Fellow}


\begin{abstract}
  We show high resolution spectra of the eclipsing brown dwarf binary
  2MASS~J05352184$-$0546085 taken at the two opposite radial velocity
  maxima. Comparisons of the TiO bands to model and template spectra
  are fully consistent with the temperatures previously derived for
  this system.  In particular, the reversal of temperatures with mass
  -- in which the higher-mass primary is cooler than its companion --
  is confirmed.  We measure the projected rotation velocities of the
  compononents; the primary is rotating at least twice as rapidly as
  the secondary.  At the two radial velocity maxima, H$\alpha$
  emission lines of both components stick out to either sides of the
  H$\alpha$ central wavelength, which is dominated by nebula emission.
  This enables us to model the individual H$\alpha$ lines of the
  primary and the secondary. We find that the H$\alpha$ emission from
  the primary is at least 7 times stronger than the emission from the
  secondary. We conclude that the temperature reversal is very likely
  due to strong magnetic fields inhibiting convection on the primary.
\end{abstract}





\section{Introduction}

2MASS~J05352184$-$0546085 (hereafter \twom) is the first known
eclipsing binary system comprising two brown dwarfs \citep{Stassun06},
with dynamically measured masses of $M_1 = 56\pm 4$ M$_{\rm Jup}$ and
$M_2 = 36\pm 3$ M$_{\rm Jup}$. The large component radii of $R_1 =
0.67\pm 0.03$ R$_\odot$ and $R_2 = 0.49\pm 0.02$ R$_\odot$ suggest
that the brown dwarfs are very young, and indeed the system satisfies
kinematic and distance criteria for membership in the Orion Nebula
Cluster (M42, age $\sim 1$ Myr). Providing the only direct and precise
measurements to date of the fundamental physical properties of young
brown dwarfs, \twom\ represents an important opportunity to test and
calibrate theoretical models of brown dwarf formation and early
evolution.

Remarkably, \twom\ exhibits a reversal of effective temperatures: The
lower-mass secondary is warmer than the higher-mass primary.  The
ratio of effective temperatures, $T_2 / T_1$, is $1.064\pm 0.004$
\citep[][hereafter SMV07]{Stassun07}. Such a temperature reversal is
not predicted by any current models of brown dwarf evolution.

One possible explanation for this temperature reversal considered by
SMV07 is that convection in the primary has been somehow
suppressed, perhaps through the presence of a strong surface field on the
primary brown dwarf. In this Letter we present high-resolution spectra of
\twom\ which (a) provide independent confirmation of the temperature reversal,
and (b) reveal that the primary is indeed much more magnetically active
than the secondary.

\section{Data}
\label{sect:data}


Data were obtained with the UVES spectrograph at ESO/VLT in March and
April, 2006 (proposal ID 276.C-5054). Two exposures of \twom\ were
taken close to the two radial velocity maxima. We give in
Table\,\ref{tab:observations} the date of observation, orbital phase,
radial velocities, and seeing conditions at the end of each exposure.
The radial velocity curve for the orbital solution of \twom\ can be
found in SMV07. We did not compute a new orbital solution because the
one given in SMV07 is already well constrained and the orbit is well
covered. Hence we do not expect significant improvements from the new
data.

Observations were taken with the slit opened to 1.1\arcsec ($R \approx
36,000$) and the CCD was used in 2x2 binning mode. Exposure times were
3600\,s. UVES was used in dichroic mode with the blue arm centered at
437\,nm and the red arm centered at 760\,nm providing wavelength
coverage from 3730\,\AA\ to 9460\,\AA\ with only a few gaps.  Because
of the target's faintness and red color, however, a low
signal-to-noise ratio (SNR) is reached below 7000\,\AA. Redward of
7000\,\AA, the SNR is around 10 per pixel.

Data were corrected for cosmic rays, bias- and dark-subtracted and
flatfielded in standard fashion using echelle reduction routines based
on the MIDAS UVES pipeline package. Sky emission was removed by
subtracting the mean of two sky spectra extracted above and below the
target spectrum in direction perpendicular to dispersion. After
3600\,s of exposure the night sky provides a significant contribution
to the spectrum in the form of sky emission lines. These lines are
mainly due to OH \citep{Osterbrock96, Hanuschik03} and can generally
be removed easily. Furthermore, \twom\ sits in the Orion nebula that
imprints a rich spectrum of strong emission lines \citep{Esteban04}.
The spectral regions of the strongest nebula emission lines are
entirely dominated by nebula lines, so that the much weaker
contribution of the target spectrum cannot be restored in the centre
of the nebula emission lines.  This is particularly problematic at the
two hydrogen lines contained in the spectra, H$\alpha$ and H$\beta$.

\section{Temperatures}


SMV07 found that the two components of \twom\ have very similar
spectral types of $\sim$M$6.5 \pm 0.5$, implying $T_\mathrm{eff}
\approx 2700$\,K. The ratio of the two components is fixed from the
ratio of the eclipses, $T_2/T_1 = 1.064 \pm 0.004$.  Absolute
temperatures could only be derived from spectral types implying
uncertainties on the order of $\sim 200$\,K.

The two TiO bands at 7050\,\AA\ ($\gamma$-band) and 8430\,\AA\
($\epsilon$-band) are particularly temperature sensitive (but gravity
insensitive). At the times of radial velocity maxima, both bandheads
can be disentangled allowing us to model the spectra of both
components. We tried to determine individual temperatures by comparing
the data at highest radial velocity difference (spectrum\,\#2 in
Table\,\ref{tab:observations}) to the sum of two reference spectra.
For the reference spectra we tried both synthetic model spectra and
empirical template spectra (see below). In both cases we assume the
radial velocity difference from the orbital solution of
SMV07.

First, we try to determine the temperatures from PHOENIX models of the
DUSTY series \cite[see][]{Allard01}. \cite{Reiners05} has shown that
the $\gamma$- and the $\epsilon$-bands yield significantly different
results in early M stars, a discrepancy probably due to uncertain
oscillator strengths mainly in the $\epsilon$-band. While temperatures
derived from the $\gamma$-band are consistent with temperatures in
M1--M5.5 stars with interferometrically measured radii, results for
the $\epsilon$-band temperatures are too low. Hence we use the
synthetic spectra only in the $\gamma$-band.  The synthetic spectra
are consistent with the temperature ratio derived by SMV07. The best
fit is achieved with $T_\mathrm{eff,1}=2850$\,K and
$T_\mathrm{eff,1}=3000$\,K.
On the other hand, these absolute temperatures are a little high
compared to the results from the $JHK$ magnitudes (SMV07).  This is
probably due to a temperature overestimate of the model spectra; in
this temperature regime, the model TiO $\gamma$-bands differ from
observed spectra in the sense that observed TiO bands show a reversal
of depth with temperature around 2700\,K. This dust-induced effect is
not fully captured by the models we used \citep{Mohanty04}; in the
model spectra the $\gamma$ bands just keep saturating.  The
$\gamma$-band has lost its high diagnostic capabilities at such low
temperatures.  Thus, while our best fit yields temperatures consistent
with the temperature reversal found by SMV07, in truth such a small
temperature difference as in the two components of \twom\ cannot be
robustly tested in our spectra of the $\gamma$-band.

\begin{figure*}
  \plotone{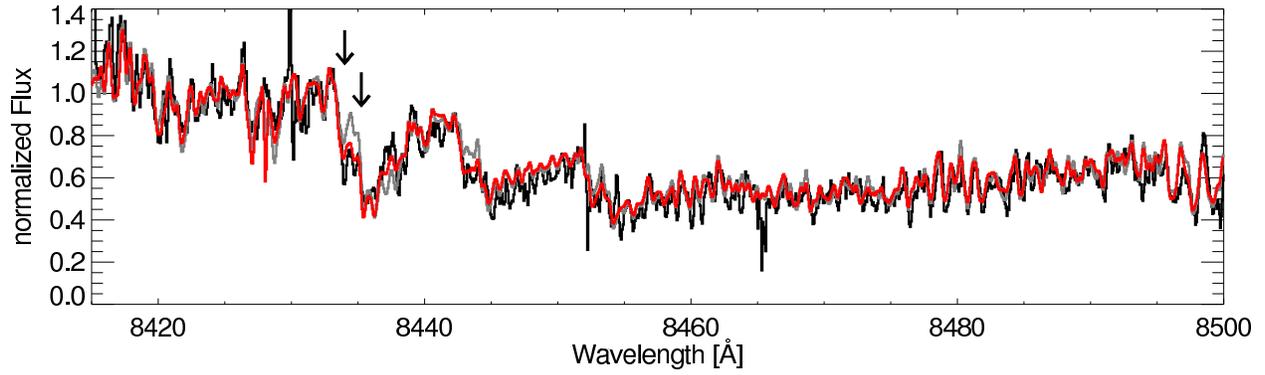}
  \caption{\label{fig:TiO} Absorption band of TiO ($\epsilon$-band).
    Data are shown in black. The two fits are the sum of two observed
    template spectra. The red line is calculated with a lower
    temperature for the primary compont, the grey line shows the
    spectrum for reversed temperatures (see text).}
\end{figure*}

In the TiO $\epsilon$-band, to circumvent known problems with the
PHOENIX models \citep{Reiners05}, we compare our data to template
spectra of single M stars observed during the same campaign. Template
stars observed are LHS\,292 (M6.5, $\sim2750$\,K) and GJ\,3877
(LHS\,3003, M7, $\sim2650$\,K).  Temperatures are from
\citet{Golimowski04}. We note that these temperatures assume a certain
radius and very high age for the template stars.  The temperatures
could be lower by about 300\,K if the template stars were young.  We
took the luminosity ratio and radial velocities from the orbital
solution as above. We show the $\epsilon$-band of \twom\ in
Fig.\,\ref{fig:TiO} (black). The red line shows our best fit, a sum of
the weighted spectra of LHS\,292 for the hotter component and of
GJ\,3877 for the cooler component.  This combination yields a very
good match to the spectrum of \twom.  Hence our data is consistent
with temperatures 2650\,K and 2750\,K for the two components in good
agreement with $JHK$ temperatures. We can also test the hypothesis
that the primary is hotter than the secondary, which is a possible
alternative from the fit to radial velocities and the photometric
light curve if the radii are reversed instead of the temperatures
(SMV07).  We show the result as a grey line in Fig.\,\ref{fig:TiO}.
Clearly, this alternative yields strong deviations from the data
around 8435\,\AA\ and 8443\,\AA. Formally, the fit qualities of the
red and the grey models in Fig.\,\ref{fig:TiO} are different by
$\Delta \chi^2 = \chi^2_{\rm red} - \chi^2_{\rm grey} = 5.5$ (for a
SNR of 10 and a reduced $\chi^2_\nu = 1.16$).  This means that the
difference is significant on a 2$\sigma$-level.

\section{Rotation and Activity}

We measure the projected rotation velocities $v\,\sin{i}$ via cross
correlation with the spectrum of GJ\,1057 (M5) observed during the
same campaign \citep[cp.][]{Reiners06}.  Several orders are used and
we consistently find the primary to be more rapidly rotating than the
secondary. We estimate our uncertainties to be 5\,km\,s$^{-1}$ due to
low SNR and the fact that two objects are in the spectra.  Our results
are given in Table\,\ref{tab:results}; we find
$v\,\sin{i}=10$\,km\,s$^{-1}$ for the primary and do not detect
significant rotation in the secondary, i.e. $v\,\sin{i} <
5$\,km\,s$^{-1}$.

\begin{figure}
\vspace{-2cm}
  \plotone{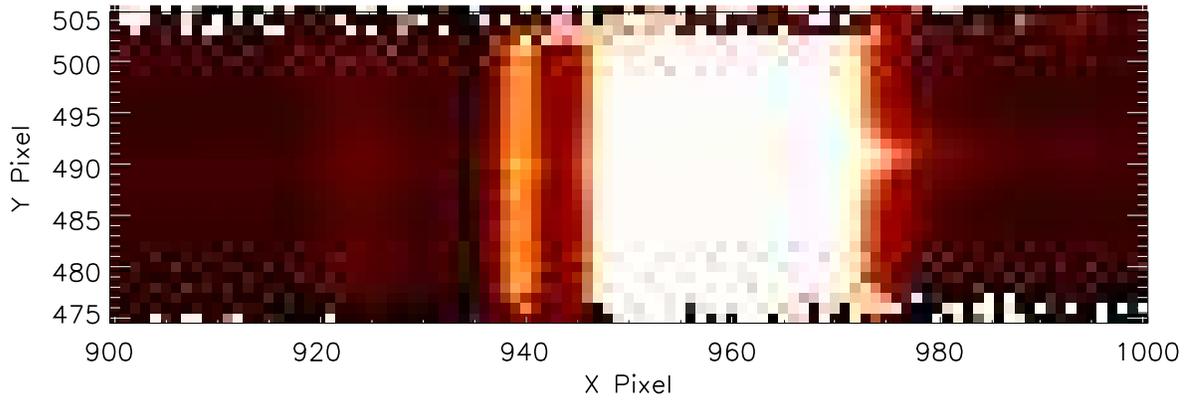}\\[-2.7cm]
  \plotone{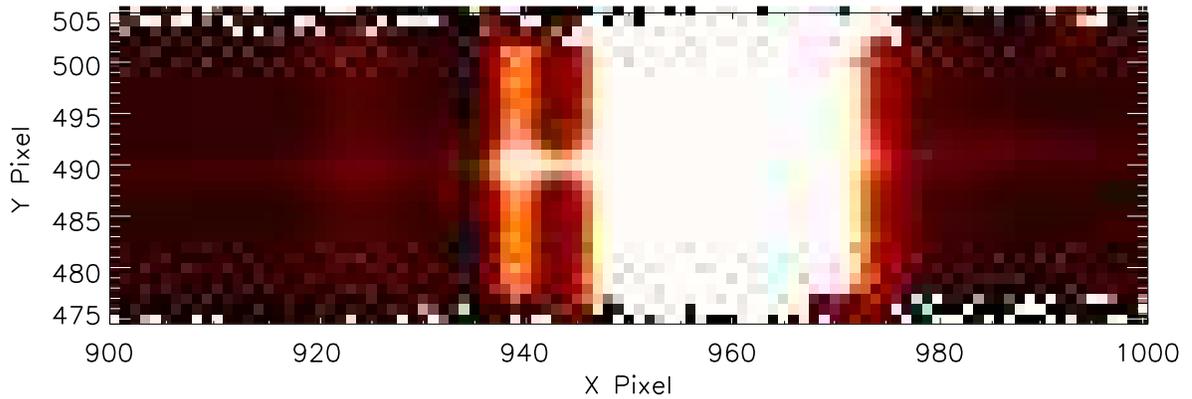}
  \caption{\label{fig:Halpha_raw} Flatfielded, cosmic-ray corrected
    raw image of the two spectra. Slit direction is vertical. The
    spectrum of \twom\ is visible around pixel 490 on the Y-axis.
    Features extended on the vertical axis are due to deuterium and
    H$\alpha$ emission of the nebula, and to the night sky emission
    (Pixel 940). H$\alpha$ emission of \twom\ appears on different
    sides of the nebula line in the two exposures. }
\end{figure}

Indicators of magnetic activity are emission in the Ca\textsc{II}~H \&
K lines (3934\,\AA\ and 3968\,\AA), or in the Ca triplet lines at
8498\,\AA, 8542\,\AA, and 8662\,\AA. All five Ca lines are covered by
our spectra, but typical Ca emission in active M dwarfs is relatively
low (unless observed during a flare). In our spectra, we do not detect
significant emission in any of the Ca lines. The strongest indicator
of magnetic activity in M-type stars is emission in H$\alpha$. This
line, however, is also emitted in H\textsc{ii} regions like the Orion
nebula in which \twom\ is located.  Flatfielded and cosmic-ray
corrected echellograms at the region of H$\alpha$ are shown in
Fig.\,\ref{fig:Halpha_raw}.  The spectrum of \twom\ appears
horizontally at pixel 490 on the Y-axis.  Around the region of
H$\alpha$, we identified three emission lines that do not come from
\twom. They are spatially extended and can be subtracted using the
area of the slit adjacent to the target spectrum.  The components are:
a) the strong H$\alpha$ nebula line, b) the weak deuterium nebula line
\citep{Hebrard00}, and c) the night sky emission line of H$\alpha$. In
these features, we see no variation between the two exposures.
Because \twom\ is comoving with M42, the two brown dwarfs produce
spectral lines fluctuating around the spectral lines of M42.  Hence
the central part of \twom's H$\alpha$ emission is obscured by the
massive nebula lines, but during the two radial velocity maxima the
H$\alpha$ lines of the two components move close to the outer edge of
the nebula emission line, and they clearly appear in the raw image in
Fig.\,\ref{fig:Halpha_raw}.

\begin{figure*}
  \plotone{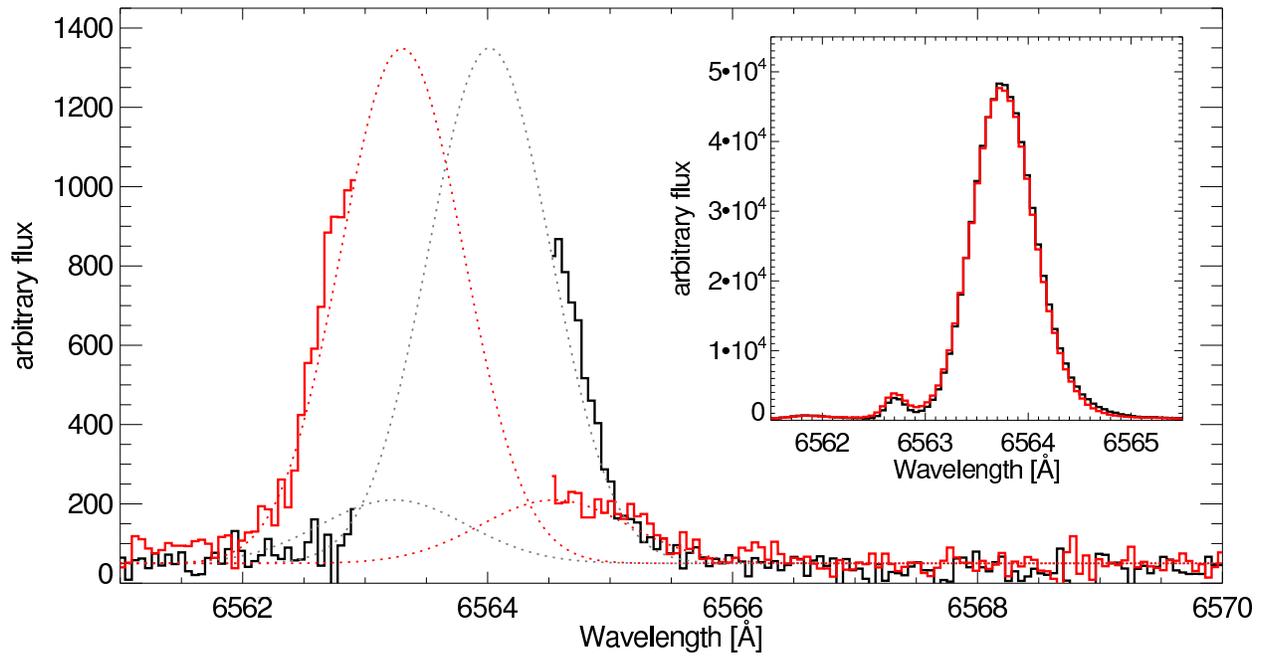}
  \caption{\label{fig:Halpha} Region around the H$\alpha$ line. Solid
    lines are spectra\,\#1 (black) and \#2 (red) with nebula and sky
    emission removed. The inset shows the original data before the
    removal. We fit two Gaussians to each spectrum (see text).
    Individual components are plotted as dotted lines, different
    colors correspond to the different spectra.}
\end{figure*}

In Fig.\,\ref{fig:Halpha}, we show the two spectra of \twom\ after
subtracting the nebular and the night sky emission lines. The black
line shows spectrum\,\#1 and the red line shows spectrum\,\#2 of
Table\,\ref{tab:observations}. In the inset, we show the reduced
spectra without subtraction of the background emission. Note that the
units are the same as in the inset, i.e. that the nebula line emission
peak is roughly a factor 50 stronger than the object's emission that
we detect in the residuals. Hence the absolute photon noise of the
nebula line emission is on the order of 200 and renders a proper
analysis in the central region of the nebula emission impossible.

It is clear from Figs.\,\ref{fig:Halpha_raw} and \ref{fig:Halpha} that
strong H$\alpha$ emission is emitted by \twom, and that its main
component is shifted to the red in spectrum\,\#1 (black line) and to
the blue in spectrum\,\#2 (red line). The data show that the primary
emits very strong H$\alpha$ emission. To quantify H$\alpha$ emission
in both components, we attempted a two component fit to the data. We
assume a Gaussian shape of the H$\alpha$ emission profiles for both
components, which is justified because we see no evidence for
accretion. We keep the positions of the two Gaussians fixed at the
wavelengths according to their radial velocities. We assume that in
both exposures the strength of each emission component is constant,
i.e. the individual equivalent widths do not vary between the two
exposures. The only free parameters in our fit are the amplitude and
the width of the Gaussians. However, because we do not see the maxima
of the emission lines, amplitude and width particularly of the primary
component's emission are degenerate.

A fit to the data is overplotted in Fig.\,\ref{fig:Halpha}, dotted
lines show the individual emission from the two components.  Black and
red color corresponds to spectra\,\#1 and \#2, respectively (same
color as the data). In spectrum\,\#1, we see only the redshifted
emission flank from the primary. No significant emission is found on
the blue side of the gap. In the second spectrum, however, we see the
blueshifted emission from the primary and a small emission component
on the red side.

Our fit shows the solution with the lowest possible emission contrast
between the two components, i.e. smallest emission from the primary
and largest emission from the secondary. For this case, we measure
individual H$\alpha$ equivalent widths of 32.6\,\AA\ for the primary
and 4.8\,\AA\ for the secondary.  One could easily fit the primary
emission using a higher amplitude for the Gaussian emission at smaller
width implying stronger primary emission. On the other hand, the
secondary emission is relatively well constrained from spectrum\,\#2,
and from spectrum\,\#1 it looks like our result is even a little on
the high end.  Thus, our measurements of H$\alpha$ equivalent widths
should be interpreted as a lower limit for the primary of \twom, and
as an upper limit for the secondary. The H$\alpha$ equivalent width of
the primary is at least a factor 7 larger than the emission from the
secondary.

\section{Conclusions}

We have used high resolution spectra of \twom\ taken at radial
velocity maxima to investigate the temperature reversal of this
enigmatic brown dwarf binary. We used two TiO bands to constrain the
absolute temperatures of the two components. In the $\gamma$-band, we
have used PHOENIX model spectra, but it is not clear to what extent
these models can accurately reproduce the real TiO spectra. From the
model spectra, we find temperatures of 2850\,K and 3000\,K ($\pm
120$\,K) for the primary and secondary, respectively. In the
$\epsilon$-band, we have used spectra from stellar templates yielding
much lower temperatures, but effects of metallicity may introduce
differences between template spectra and the spectrum of \twom. This
method yields temperatures between 2600\,K and 2750\,K.  Most
importantly, the $\epsilon$-band analysis confirms the finding by
SMV07 of a temperature reversal in this system, such that the
higher-mass primary is indeed cooler than the lower-mass secondary.

A possible explanation for the temperature reversal may be that large
scale magnetic fields are inhibiting convection. In low mass stars,
magnetic fields manifest themselves for example as H$\alpha$ emission,
and at least in sun-like stars activity is correlated to the rotation
rate. We find that the primary of \twom\ shows at least 7 times
stronger activity than the secondary, and the former is also rotating
at least a factor of 2 more rapidly than the latter.

We can convert the equivalent widths into normalized H$\alpha$
luminosities \cite[see][]{Reiners07}. We find $\log{L_\textrm{\small
    H$\alpha$}/L_\textrm{\small bol}} = -3.47$ and $-4.30$ for the
primary and secondary components, respectively.  \citet{Reiners07}
have measured H$\alpha$ luminosities and magnetic fields in a sample
of M dwarfs. They found that for a given spectral type H$\alpha$ is
closely correlated with magnetic field. From the normalized H$\alpha$
luminosities of \twom\ we can estimate the mean magnetic flux on its
two componenents. The primary's H$\alpha$ luminosity of
$\log{L_\textrm{\small H$\alpha$}/L_\textrm{\small bol}} = -3.47$ is
on the top end of H$\alpha$ luminosities found among M6--M7 dwarfs and
indicate a very high magnetic flux level of $Bf \approx 4$\,kG over
the entire object \citep[cp. left panel of Fig.\,12 in][]{Reiners07}.
The secondary's H$\alpha$ luminosity is only around average for that
spectral range \citep[see also][]{West04} and indicates significantly
weaker magnetic flux of $Bf \approx 2$\,kG.

Recently, \citet{Chabrier07} have shown that the effect of inhibited
convection could indeed explain the temperature reversal in \twom.
They suggest that magnetic fields with surface values of a few kG
might be sufficient for severely inhibiting convection in parts of low
mass star interiors. 
The observations presented here support this scenario.

\acknowledgements

We thank Michael Sterzik, it was his idea to obtain such a data set.
Based on observations collected at the European Southern Observatory,
Paranal, Chile, 276.C-5054.  A.R. has received research funding from
the DFG as an Emmy Noether fellow (RE 1664/4-1). A.S. acknowledges
funding from the DFG (RE 1664/4-1). K.G.S. gratefully acknowledges
support from National Science Foundation Career grant AST-0349075 and
from a Research Corporation Cottrell Scholar award.  R.D.M.
acknowledges NSF Grant AST-0406615.





\clearpage

\clearpage

\begin{deluxetable}{ccccccc}
  \tablecaption{\label{tab:observations} UVES observations of
    2M0535$-$05.}  \tablewidth{0pt}
  \tablehead{\# & JD\tablenotemark{a} - 2\,453\,800 & Orbital phase & Seeing & \multicolumn{2}{c}{$v_\textrm{\tiny rad}${\tiny [km\,s$^{-1}$]}}\\
    &&&& {\tiny prim} & {\tiny sec}} \startdata
  1 & 31.49344 & 0.2445--0.2487 & 1.40\arcsec & $+ 12.6$ & $-20.0$\\
  2 & 28.49346 & 0.9377--0.9420 & 1.25\arcsec & $- 23.2$ & $+36.7$
  \enddata
  \tablenotetext{a}{Julian Date at start of observation}
\end{deluxetable}

\clearpage

\begin{deluxetable}{lcc}
  \tablecaption{\label{tab:results} Rotation and activity of \twom.}
  \tablewidth{0pt}
  \tablehead{ & $v\,\sin{i}$ [km\,s$^{-1}$] & $\log{L_\textrm{\tiny H$\alpha$}/L_\textrm{\tiny bol}}$}
  \startdata
  Primary   &   10 & $-3.47$\\
  Secondary & $<5$ & $-4.30$
  \enddata
\end{deluxetable}

\clearpage

\end{document}